\documentstyle{article}[12pt]
\newcommand{\nin}{\noindent}
\newcommand{\be}{\begin{equation}}
\newcommand{\ee}{\end{equation}}
\newcommand{\bea}{\begin{eqnarray}}
\newcommand{\eea}{\end{eqnarray}}

\newcommand{\hf}{\frac{1}{2}}
\newcommand{\nonu}{\nonumber\\}

\newcommand{\ol}{\overline}

\begin{document}

\begin{center}

{\Large{\bf A control on quantum fluctuations\\ in 2+1 dimensions}}

\vspace{1cm}

{\bf Jean Alexandre}\\
Physics Department, King's College \\
WC2R 2LS, London, UK\\
jean.alexandre@kcl.ac.uk

\vspace{3cm}

{\bf Abstract}

\end{center}

\vspace{.2cm}

\nin A functional method is discussed, where the quantum fluctuations of a theory are controlled by
a mass parameter and the evolution of the theory with this parameter is connected to 
its renormalization. It is found, in the framework of the gradient
expansion, that the coupling constant of a $N=1$ Wess-Zumino theory in 2+1 dimensions
does not get quantum corrections. 

\vspace{3cm}

\section{Introduction}

The understanding of the notion of renormalization had done a great step 
forward when the connection with the blocking procedure was made
\cite{wilson}. In this framework, the control
of the quantum fluctuations can be performed by the progressive elimination
of the Fourier components of the fields in a given theory \cite{block}. The parameters of the 
theory are then functions of the running cut-off and the corresponding 
renormalization flows describe explicitly the dependence of the theory on 
the energy scale. There are different approaches here, due to the freedom in 
defining the coarse-graining procedure. 

So as to avoid the dependence on a specific blocking procedure,
an alternative approach to renormalization was proposed \cite{functscal}.
There the quantum fluctuations are controlled by the mass of a scalar field: 
a large mass freezes the quantum corrections, the system is classical, and as the 
mass decreases the fluctuations appear and generate the quantum effects.
The evolution of the parameters with this mass can thus be seen as renormalization flows
and it was shown that these correspond to the usual flows at the one-loop level.
Beyond one-loop, the "mass-controlled" flows do not coincide anymore with the loop
expansion since they are generated in the framework of the gradient expansion, which 
is based on an expansion of the effective action (proper graphs generating functional) in
derivatives of the field, rather than based on an expansion in powers of $\hbar$. 
An example of the difference with a loop expansion is given in \cite{functqed}, where
the same method is applied to QED and the running mass is the fermions one.
There the famous Landau pole is recovered
at one-loop, but disappears if all the quantum corrections are taken into account. It 
should not be forgotten that this result holds at the lowest order in the gradient 
expansion and higher order derivatives were not taken into account.
Finally, this functional method was also applied to the study of the Coulomb interaction
in an electron gas \cite{functcondmat} where the one-loop Lindhard function and screening
are reproduced.

In the present paper, we propose to make use of this functional method in 2+1 dimensions, 
motivated by relativistic-like effective theories for the description of high-temperature
planar superconductors \cite{highTc}. The use of renormalization procedures here is 
essential: starting from lattice models dealing with the initial strongly correlated electron 
system, the corresponding theory in the continuum should give to the fore the relevant 
operators and interactions, and the present method is believed to be well appropriate for this.

Section 2 presents the method in the simple case of a self-interacting scalar field, so 
as to concentrate on the physical meaning of the flows that are obtained. The connection 
with the usual renormalization flows is made and the technical details, as well as the
derivation of the evolution equation, are given in the Appendix A. 

Section 3 deals with the supersymmetric generalization of the previous
case. $N=1$ supersymmetry in 2+1 dimensions does not have non-renormalization
theorems, since the integrations over superspace have all the same measure $d^3xd^2\theta$
and no chiral superfield can be defined.
It is thus interesting to make use of this functional method to learn about the 
behaviour of the parameters under the influence of quantum corrections. It will be
found that, to the order of the gradient expansion and the truncation of the potential which 
are considered, the 
coupling constant does not get quantum corrections, at any order in $\hbar$. 
This absence of renormalization of the parameters, though, should not be confused with
a non-renormalization theorem: higher powers of the superfield or higher order terms in 
the gradient expansion (higher derivatives, derivative interactions) would lead to
a renormalization of the coupling constant.
For the sake of clarity in the presentation, the derivations are given in the Appendix B. 

Finally, the conclusion contains a discussion on the gradient expansion approximation.

\section{Self interacting scalar field}

We present here the functional method in the case of a self interacting scalar field.
The difference with 3+1 dimensions is that the integrals do not need regularization, 
as will be seen.

The starting point is the Euclidean action

\be
S=\int d^3x \left\{-\hf\phi\Box\phi+\frac{\lambda}{2}m_0^2\phi^2+\frac{e_0}{24}\phi^4\right\},
\ee

\nin where the mass dimension of the coupling is $[e_0]=1$.
The classical system will
correspond to $\lambda\to\infty$, where the mass is infinite and thus the quantum effects
are not present. These will gradually appear as $\lambda$ decreases, and when $\lambda\to 1$
the full quantum corrections will be present.

As usual, we introduce the effective action $\Gamma$
as the Legendre transform of the connected graphs generator functional. 
After some manipulations (see Appendix A), we find then the following exact evolution 
equation with $\lambda$: 

\be\label{evolfinale}
\partial_\lambda\Gamma=
\frac{m_0^2}{2}\int d^3x\left\{\phi^2(x)+
\left(\frac{\delta^2\Gamma}{\delta\phi^2(x)}\right)^{-1}\right\}
\ee

\nin To obtain informations on the physical processes, 
we express the functional $\Gamma$ in terms of its argument $\phi$ via a gradient 
expansion and a truncation of the potential, to assume the following functional form:

\be
\Gamma=\int d^3x \left\{-\hf\phi\Box\phi+\frac{\lambda}{2}m^2\phi^2+\frac{e}{24}\phi^4\right\}.
\ee

\nin In this approximation, we obtain the following evolution of the parameters
(see Appendix A):

\bea\label{equadiffinal}
\partial_\lambda(\lambda m^2)&=&m_0^2\left(1-\frac{e}{16\pi\lambda^{1/2}m}\right)\nonu
\partial_\lambda e&=&\frac{3e^2m_0^2}{32\pi\lambda^{3/2}m^3}.
\eea

One should bare in mind that the equations (\ref{equadiffinal}) contain all the quantum
corrections, in this approximation of the gradient expansion. The connection with the
one-loop results is obtained by making the replacements $m^2\to m_0^2$ and $e\to e_0$
in the right hand sides of Eqs.(\ref{equadiffinal}), since restoring the factors $\hbar$
leads to $e\to\hbar e$:

\bea
\partial_\lambda(\lambda m^2)&=&m_0^2-\hbar\frac{e_0m_0}{16\pi\lambda^{1/2}}
+{\cal O}(\hbar^2)\nonu
\partial_\lambda e&=&\hbar\frac{3e_0^2}{32\pi\lambda^{3/2}m_0}
+{\cal O}(\hbar^2).
\eea

\nin The solutions are found by integrating from $\lambda=\infty$ to $\lambda=1$:

\bea\label{oneloopint}
m^2-m_0^2&=&-M^2+\hbar\frac{e_0m_0}{8\pi}+{\cal O}(\hbar^2)\nonu
e-E&=&-\hbar\frac{3e_0^2}{16\pi m_0}+{\cal O}(\hbar^2),
\eea

\nin where $M^2$ and $E$ are constant of integration.
The solutions (\ref{oneloopint}) coincide with the usual one-loop results
obtained when computing the appropriate Feynman graphs (see Appendix A), where 
the scale $M$ is linked to the cut-off of the theory and $E=e_0$.
Therefore it has been shown that the evolution with the parameter $\lambda$ 
controls the quantum fluctuations and is consistent, at one loop,  with the
usual renormalization scheme.

\section{$N=1$ Wess-Zumino model}

We deal here with the supersymmetric generalization of the previous example. 

In terms of the real scalar superfield $Q$,
the supersymmetric generalization of the previous section is the Wess-Zumino model,
with the following classical Euclidean action 

\be
S[Q]=\int d^5z\left\{\hf Q D^2 Q+\frac{\lambda}{2}m_0 Q^2
+\frac{g_0}{6} Q^3\right\},
\ee

\nin where $z=(x,\theta)$ is the superspace coordinate 
and the conventions are those taken in \cite{gates}. 
The coupling $g_0$ has mass dimension $[g_0]=1/2$ and the on-shell Lagrangian
contains, among other terms, the interaction $g_0^2\phi^4$, where $\phi$ is the 
scalar component of $Q$.

The evolution equation of the proper graphs generator functional $\Gamma$ has a similar 
form as the one obtained in the non-supersymmetric case and is

\be\label{evolsusy}
\partial_\lambda\Gamma=\frac{m_0}{2}\int d^5z\left\{Q^2(z)
+\left(\frac{\delta^2\Gamma}{\delta Q^2(z)}\right)^{-1}\right\}
\ee

\nin In the framework of the gradient expansion, and truncating the potential to 
the order $Q^3$, we consider the following ansatz for the 
functional dependence of $\Gamma$:

\be\label{gradexpsusy}
\Gamma[Q]=\int d^5z\left\{\hf Q D^2 Q+\frac{\lambda}{2}m Q^2
+\frac{g}{6}Q^3\right\},
\ee

\nin In this approximation, the trace of the operator $(\delta^2\Gamma)^{-1}$
is computed in the Appendix B and it is shown that the relevant quadratic term,
as far as the evolution of the mass is concerned, is

\bea\label{evolmk}
&&g^2\int\frac{d^3k}{(2\pi)^3}d^2\theta ~~ Q(k,\theta)Q(-k,\theta)\\
&&\times\int\frac{d^3p}{(2\pi)^3}
\frac{[p^2-3(\lambda m)^2]}{[p^2+(\lambda m)^2]^2[(k-p)^2+(\lambda m)^2]}.\nonumber
\eea

\nin In the approximation (\ref{gradexpsusy}), 
the quantum corrections to the mass are obtained when $k=0$ in the integrand.
The identification of the non-derivative quadratic terms in 
both sides of the evolution equation (\ref{evolsusy}) gives then

\be\label{evolm}
\partial_\lambda(\lambda m)=m_0+g^2m_0\int\frac{d^3p}{(2\pi)^3}
\frac{p^2-3(\lambda m)^2}{[p^2+(\lambda m)^2]^3}.
\ee

\nin The integration over $p$ is easy and gives a vanishing integral, such that

\be
\partial_\lambda(\lambda m)=m_0,
\ee

\nin i.e. the evolution of $(\lambda m)$ is classical and does not have quantum corrections. 
The evolution of $m$ with $\lambda$ is then 

\be\label{mfinal}
m=m_0\left(1+\frac{c}{\lambda}\right),
\ee

\nin where $c$ is a dimensionless constant, which is computed at the one-loop level 
in the Appendix B, using the Feynman graph technique. We can note that in a 
numerical resolution of the differential equation for the mass, the 
large but {\it finite} initial value for $\lambda$ would not allow us to set exactly
$m=m_0$ since quantum fluctuations, even tiny, are present for finite $\lambda$.
This is the reason why the constant of integration $c$ would not vanish.

Let us look at the evolution of the coupling constant $g(\lambda)$.
It is seen in the Appendix B that the relevant cubic term, as far as the evolution
of the coupling is concerned, is

\bea
&&g^3\int\frac{d^3q}{(2\pi)^3}\frac{d^3k}{(2\pi)^3}d^2\theta
~~ Q(k,\theta)Q(q,\theta)Q(-k-q,\theta)\\
&&\times\int\frac{d^3p}{(2\pi)^3}\frac{4(\lambda m)[(\lambda m)^2-p^2]}
{[p^2+(\lambda m)^2]^2[(p+q)^2+(\lambda m)^2][(p-k)^2+(\lambda m)^2]}.\nonumber
\eea

\nin In the framework of the approximation (\ref{gradexpsusy}),
the quantum corrections to the coupling are obtained when $k=q=0$ in the integrand.
The identification of the non-derivative cubic terms in both sides of the evolution equation 
(\ref{evolsusy}) gives then

\be\label{evolg}
\partial_\lambda g=12g^3(\lambda m) m_0\int\frac{d^3p}{(2\pi)^3}
\frac{(\lambda m)^2-p^2}{[p^2+(\lambda m)^2]^4}.
\ee

\nin Once again, the integration over $p$ is easy and gives a vanishing integral,
such that

\be\label{finalg}
\partial_\lambda g=0,
\ee

\nin i.e. the coupling constant does not evolve with $\lambda$. This is consistent
with the one-loop result (see Appendix B) and provides a generalization to all 
orders in $\hbar$, in the approximation (\ref{gradexpsusy}).

Finally, the quantities 
$\lambda m$ and $g$ are frozen to their initial value given when $\lambda\to\infty$, i.e. when the 
quantum fluctuations can be neglected. The coupling constant $g$, in the present
approximation (\ref{gradexpsusy}), does not get any quantum correction.

\section{Conclusion}

To conclude, let us stress again the non-perturbative nature of the flows that are obtained with 
this method. The evolution equations of the coupling constants contain all 
the quantum corrections, independently of any pertubative expansion in $\hbar$, 
{\it in the approximation of the gradient expansion and the polynomial truncation of the
potential}. This feature is 
interesting in the context of low-energy effective theories, where only low 
powers of the momentum are kept into account in the functional dependence of the effective
action, as well as relevant operators only, in a renormalization group sense.

Let us be more precise concerning this gradient expansion approximation. 
The evolution equation of the effective action, giving $\partial_\lambda\Gamma$,
is exact and contains all the quantum corrections. 
However, it looks like a one-loop equation, since it has the form (when the 
factor $\hbar$ is restored)  

\be\label{evolgene}
\partial_\lambda\Gamma=\hbar{\cal F}[\Gamma],
\ee

\nin where ${\cal F}$ is some operator applied on $\Gamma$. This is the reason why the 
usual one-loop results are
obtained, when $\Gamma$ is replaced by the classical action $S$ in the right-hand side of Eq.(\ref{evolgene}).
The next step forward, beyond one-loop, is to assume a functional dependence of $\Gamma$ and to plug it inside
the evolution equation (\ref{evolgene}). The resulting couplings' evolution consist then in a partial 
resumation of the graphs. In the present example, where no new vertices were considered compared to the
classical ones, the resumation takes into account the one-loop-like graphs, with the internal lines being 
the full propagators (full in the present approximation of the gradient expansion and polynomial 
truncation of the potential). 

Note that more general gradient expansions could be considered, besides
including higher order powers of the field: one could add
higher order derivatives in the quadratic kinetic term, as well as
derivative interactions. These would lead to new vertices and therefore to a more complete resumation
of the graphs.

Finally, let us make a comment on $N=2$ supersymmetry. The latter can be obtained by
dimensional reduction of $N=1$ supersymmetry in 3+1 dimensions and thus
has non-renormalization theorems. The present method can then still be useful
for the study of the kinetic operators, since these do get renormalized. 
The evolution equation with the parameter $\lambda$ would be quite different though,
due to the existence of two kinds of integrals: $\int d^3x d^2\theta$ and 
$\int d^3x d^2\theta d^2\ol\theta$. Indeed, the functional derivatives have to be 
taken with respect to the full superspace dependence, such that \cite{gates}

\be
\frac{\delta Q(x,\theta,\ol\theta)}{\delta Q(x',\theta',\ol\theta')}=
\ol D^2\delta^2(\theta-\theta')\delta^2(\ol\theta-\ol\theta')\delta^3(x-x')
\ee

\nin when $Q$ is a chiral superfield. As a consequence, the 
final evolution equation is more involved and is planned for a future work.

\section*{Acknowledgments}

I would like to thank the Physics Department of King's College for its support.

\section*{Appendix A: Evolution equation for the 
scalar field}

The Euclidean action, functional of the field $\hat\phi$, is

\be
S[\hat\phi]=\int d^3x \left\{-\hf\hat\phi\Box\hat\phi+
\frac{\lambda}{2}m_0^2\hat\phi^2+\frac{e_0}{24}\hat\phi^4\right\},
\ee

\nin and the connected graphs generator functional, function of the source $j$,
is defined by $W[j]=-\ln{\cal Z}[j]$ where

\be
{\cal Z}[j]=\int{\cal D}[\hat\phi]\exp\left\{-S[\hat\phi]-\int d^3x j(x)\hat\phi(x)\right\}.
\ee

\nin The functional derivative of $W$ defines the expectation value field $\phi$:

\be\label{defquant}
\frac{\delta W}{\delta j(x)}=\frac{1}{{\cal Z}}<\hat\phi(x)>=\phi(x),
\ee

\nin where

\be
<\hat\phi(x)>=\int{\cal D}[\hat\phi] \hat\phi(x)
\exp\left\{-S[\hat\phi]-\int d^3y j(y)\hat\phi(y)\right\}.
\ee

\nin We also have

\be\label{d2W}
\frac{\delta^2 W}{\delta j(x)\delta j(y)}
=\phi(x)\phi(y)-\frac{1}{{\cal Z}}<\hat\phi(x)\hat\phi(y)>.
\ee

\nin Inverting the relation (\ref{defquant}) which gives $\phi(x)$ as a function of $j(x)$,
we define the Legendre transform $\Gamma$ (functional of $\phi$) of $W$ by

\be
\Gamma[\phi]=W[j]-\int d^3x j(x)\phi(x).
\ee

\nin From this definition we extract the following functional derivatives:

\bea\label{d2G}
\frac{\delta\Gamma}{\delta\phi(x)}&=&-j(x)\\
\frac{\delta^2\Gamma}{\delta\phi(x)\delta\phi(y)}&=&
-\left(\frac{\delta^2 W}{\delta j(x)\delta j(y)}\right)^{-1}\nonumber
\eea

\nin The evolution of $W$ with the parameter $\lambda$ is, according to (\ref{d2W}),

\bea
\partial_\lambda W&=&-\frac{m_0^2}{2{\cal Z}}
\int d^3x<\hat\phi^2(x)>\nonu
&=&\frac{m_0^2}{2}\int d^3x\phi^2(x)
-\frac{m_0^2}{2}\int d^3x\frac{\delta^2 W}{\delta j^2(x)}.
\eea

\nin To compute the evolution of $\Gamma$ with $\lambda$, we have to keep in mind that its
independent variables are $\phi$ and $\lambda$, such that

\be
\partial_\lambda \Gamma=\partial_\lambda W+
\int d^3x \frac{\delta W}{\delta j(x)}\partial_\lambda j(x)
-\int d^3x\partial_\lambda j(x)\phi(x)=\partial_\lambda W.
\ee

\nin Combining these different results, we finally obtain the exact evolution equation for
the proper graphs generator functional $\Gamma$:

\bea\label{evol}
\partial_\lambda\Gamma&=&
\frac{m_0^2}{2}\int d^3x\left\{\phi^2(x)+
\left(\frac{\delta^2\Gamma}{\delta\phi^2(x)}\right)^{-1}\right\}\\
&=&\frac{m_0^2}{2}\int\frac{d^3p}{(2\pi)^3}\left\{\phi(p)\ol\phi(p)+
\left(\frac{\delta^2\Gamma}{\delta\phi(p)\delta\ol\phi(p)}\right)^{-1}\right\},\nonumber
\eea

\nin where we used the fact that $\phi(-p)=\ol\phi(p)$ since $\phi(x)$ is real. 
Note that, in Eq.(\ref{evol}), the integration of the operator $(\delta^2\Gamma)^{-1}$ 
has to be understood as

\be
\int\frac{d^3p}{(2\pi)^3}\frac{d^3q}{(2\pi)^3}
\left(\frac{\delta^2\Gamma}{\delta\phi(p)\delta\phi(q)}\right)^{-1}(2\pi)^3\delta^3(p+q).
\ee

In the approximation where we assume that

\be
\Gamma=\int d^3x \left\{-\hf\phi\Box\phi+\frac{\lambda}{2}m^2\phi^2+\frac{e}{24}\phi^4\right\},
\ee

\nin it is enough to consider a constant configuration $\phi_0$ of 
$\phi(x)$, which leads to

\bea\label{deriv}
\Gamma[\phi_0]&=&{\cal V}\left(\frac{\lambda}{2}m^2\phi_0^2+\frac{e}{24}\phi_0^4\right)\nonu
\left(\frac{\delta^2\Gamma}{\delta\phi(p)\delta\phi(q)}\right)^{-1}_{\phi_0}&=&
\frac{\delta^3(p+q)}{p^2+\lambda m^2+\frac{e}{2}\phi_0^2},
\eea

\nin where ${\cal V}=\delta^3(0)$ is the space-time volume. 
The evolution equations for $m^2$ and $e$ are obtained after expanding the right-hand
side of Eq.(\ref{deriv}) in powers of $\phi_0$, which leads to:

\bea
&&\int\frac{d^3p}{(2\pi)^3}
\left(\frac{\delta^2\Gamma}{\delta\phi(p)\delta\ol\phi(p)}\right)^{-1}_{\phi_0}\nonu
&=&{\cal V}\left\{\mbox{constant}
-\frac{e}{2(2\pi)^3}\left(\int\frac{d^3p}{(p^2+\lambda m^2)^2}\right)\phi_0^2\right.\nonu
&&\left.~~~~~~~~+\frac{e^2}{4(2\pi)^3}
\left(\int\frac{d^3p}{(p^2+\lambda m^2)^3}\right)\phi_0^4+...\right\}\nonu
&=&{\cal V}\left\{\mbox{constant}-\frac{e}{16\pi\lambda^{1/2}m}\phi_0^2
+\frac{e^2}{128\pi\lambda^{3/2}m^3}\phi_0^4+...\right\},
\eea

\nin such that the identification of the coefficients
of $\phi_0^2$ and $\phi_0^4$ in the evolution equation (\ref{evol}) leads to

\bea\label{equadiff}
\partial_\lambda(\lambda m^2)&=&m_0^2\left(1-\frac{e}{16\pi\lambda^{1/2}m}\right)\nonu
\partial_\lambda e&=&\frac{3e^2m_0^2}{32\pi\lambda^{3/2}m^3}.
\eea

\vspace{1cm}

We finally compute the one-loop renormalization of the potential of this model,
using the usual Feynman rules, and starting with the action for
$\lambda=1$:

\be
S_{\lambda=1}=\int d^3x \left\{-\hf\phi\Box\phi+\frac{m_0^2}{2}\phi^2+\frac{e_0}{24}\phi^4\right\}.
\ee

The one-loop correction to the mass is given by the tadpole diagram. Taking into
account the symmetry factor (1/2), we have 

\bea
m^2&=&m_0^2-\hbar\frac{e_0}{2}\int\frac{d^3q}{(2\pi)^3}\frac{1}{q^2+m^2_0}\nonu
&=&m_0^2-\hbar\frac{e_0\Lambda}{4\pi^2}+\hbar\frac{e_0m_0}{8\pi}
+{\cal O}\left(\frac{1}{\Lambda}\right)
\eea

\nin where $\Lambda$ is the cut-off. This last result coincides with Eq.(\ref{oneloopint}) 
if we make the identification $M^2\to\hbar e_0\Lambda/4\pi^2$.

The one-loop correction to the charge is given by, at the limit of zero incoming momenta and 
taking into account the symmetry factor (1/2) as well as the 3 different possibilities for the 
incoming momenta:

\bea
e&=&e_0+\hbar\frac{3}{2}(ie)^2\int\frac{d^3q}{(2\pi)^3}\frac{1}{(q^2+m^2_0)^2}\nonu
&=&e_0-\hbar\frac{3e_0^2}{16\pi m_0},
\eea

\nin which is consistent with the result (\ref{oneloopint}).

\section*{Appendix B: Evolution equation for the Wess-Zumino model}

The reader can find a complete presentation
of superspace in 2+1 dimensions in \cite{gates,hitchin} and we give here only the basic 
properties necessary for the present derivation. The conventions are those given in
\cite{gates}.

The classical action we are interested in, functional of the (real) $N=1$ 
superfield $\hat Q$, is given by

\be
S[\hat Q]=\int d^5z\left\{\hf \hat Q D^2\hat Q+\frac{\lambda}{2}m_0 \hat Q^2
+\frac{g_0}{6}\hat Q^3\right\},
\ee

\nin where $d^5z=d^3xd^2\theta$ and the spinorial derivative is

\be
D^\alpha=\frac{\partial}{\partial\theta_\alpha}+i(\gamma^\mu\theta)^\alpha\partial_\mu,
\ee

\nin where the gamma matrices are given by
$\gamma^0=\sigma^2,\gamma^1=i\sigma^1,\gamma^2=i\sigma^3$,
and $\sigma^1,\sigma^2,\sigma^3$ are the Pauli matrices.

The construction of the expectation value superfield $Q$ follows the one given in the Appendix A:
if $W$ is the connected graphs generator functional, function of the supersource $P$,
we have

\be
\frac{\partial W}{\partial P(z)}=Q(z).
\ee

\nin Following the steps of Appendix A, it is straightforward to see
that the evolution equation of the proper graphs generator functional $\Gamma$ is

\bea\label{evolsusy1}
\partial_\lambda\Gamma&=&\frac{m_0}{2}\int d^5z\left\{Q^2(z)
+\left(\frac{\delta^2\Gamma}{\delta Q^2(z)}\right)^{-1}\right\}\\
&=&\frac{m_0}{2}\int \frac{d^3p}{(2\pi)^3}d^2\theta\left\{Q(p,\theta)\ol Q(p,\theta)
+\left(\frac{\delta^2\Gamma}{\delta Q(p,\theta)\delta \ol Q(p,\theta)}\right)^{-1}\right\}\nonumber,
\eea

\nin where we used $Q(-p,\theta)=\ol Q(p,\theta)$ since $Q(z)$ is real. 
In Eq.(\ref{evolsusy1}), the integration of the operator $(\delta\Gamma)^{-1}$
has to be understood as

\be\label{trace}
\int \frac{d^3p}{(2\pi)^3}\frac{d^3q}{(2\pi)^3}d^2\theta d^2\theta'
\left(\frac{\delta^2\Gamma}{\delta Q(p,\theta)\delta Q(q,\theta')}\right)^{-1}
(2\pi)^3\delta^3(p+q)\delta^2(\theta-\theta').
\ee

\nin With the following functional dependence of $\Gamma$: 

\be\label{GammaQ}
\Gamma[Q]=\int d^5z\left\{\hf Q D^2 Q +\frac{\lambda}{2}m Q^2
+\frac{g}{6}Q^3\right\},
\ee

\nin we obtain

\bea
\frac{\delta^2\Gamma}{\delta Q(p,\theta)\delta Q(q,\theta')}&=&
(\lambda m+D^2_{p\theta})(2\pi)^3\delta^3(p+q)\delta^2(\theta-\theta')\nonu
&&+gQ(p+q,\theta)\delta^2(\theta-\theta'),
\eea

\nin where $D^\alpha_{p\theta}=\partial^\alpha-p_\mu(\gamma^\mu\theta)^\alpha$.
We now take the inverse of $(\delta^2\Gamma)$, using the expansion

\bea\label{expmat}
(A+B)^{-1}&=&A^{-1}-A^{-1}BA^{-1}+A^{-1}BA^{-1}BA^{-1}\nonu
&&-A^{-1}BA^{-1}BA^{-1}BA^{-1}+...
\eea

\nin where 

\bea
A&=&(\lambda m+D^2_{p\theta})(2\pi)^3\delta^3(p+q)\delta^2(\theta-\theta')\nonu
B&=&gQ(p+q,\theta)\delta^2(\theta-\theta'),
\eea
 
\nin such that $A$ is diagonal in Fourier space. 
We will also use the following properties \cite{gates}:

\bea\label{propgates}
&&\delta^2(\theta_1-\theta_2)\delta^2(\theta_2-\theta_1)=0\nonu
&&\delta^2(\theta_1-\theta_2)D^\alpha_{p\theta_1}\delta^2(\theta_2-\theta_1)=0\nonu
&&\delta^2(\theta_1-\theta_2)D^2_{p\theta_1}\delta^2(\theta_2-\theta_1)=\delta^2(\theta_1-\theta_2)\nonu
&&(D^2_{p\theta})^2=-p^2,
\eea

\nin such that 

\be
A^{-1}=\frac{\lambda m-D^2_{p\theta}}{p^2+(\lambda m)^2}
(2\pi)^3\delta^3(p+q)\delta^2(\theta-\theta').
\ee

\vspace{.5cm}

In the expansion (\ref{expmat}), the constant and linear terms in $B$ lead to 
constants after taking the trace (\ref{trace}).
The trace of the quadratic term is

\bea
&&g^2\int\frac{d^3p}{(2\pi)^3}\frac{d^3k}{(2\pi)^3}d^2\theta d^2\theta_1 d^2\theta_2\\ 
&&\times Q(-p+k,\theta_1) Q(-k+p,\theta_2)\nonu
&&\times
\frac{\lambda m-D^2_{p\theta}}{p^2+(\lambda m)^2}\delta^2(\theta-\theta_1)
\frac{\lambda m-D^2_{k\theta_1}}{k^2+(\lambda m)^2}\delta^2(\theta_1-\theta_2)\nonu
&&\times
\frac{\lambda m-D^2_{p\theta_2}}{p^2+(\lambda m)^2}\delta^2(\theta_2-\theta)\nonumber.
\eea

\nin Using the properties (\ref{propgates}), we obtain then for the quadratic term

\bea
&&g^2\int\frac{d^3p}{(2\pi)^3}\frac{d^3k}{(2\pi)^3}d^2\theta
\frac{1}{[p^2+(\lambda m)^2]^2[(k-p)^2+(\lambda m)^2]}\\
&&\times\left(3\lambda mQ(k,\theta)D^2_{k-p} Q(-k,\theta)+
[p^2-3(\lambda m)^2]Q(k,\theta)Q(-k,\theta)\right).\nonumber
\eea

\nin In the previous integral, both terms give a kinetic contribution,
when $k\ne 0$. In the present approximation (\ref{GammaQ}), we neglect
these kinetic terms and the remaining contribution is obtained for $k=0$ in the second integrand. 
This gives the quantum corrections to the mass, obtained from Eq.(\ref{evolmk}).

\vspace{.5cm}

The trace of the cubic term gives

\bea
&&-g^3\int\frac{d^3p}{(2\pi)^3}\frac{d^3q}{(2\pi)^3}\frac{d^3k}{(2\pi)^3}
d^2\theta d^2\theta_1 d^2\theta_2d^2\theta_3\\
&&\times Q(-p+q,\theta_1) Q(-q+k,\theta_2)Q(-k+p,\theta_3)\nonu
&&\times
\frac{\lambda m-D^2_{p\theta}}{p^2+(\lambda m)^2}\delta^2(\theta-\theta_1)
\frac{\lambda m-D^2_{q\theta_1}}{q^2+(\lambda m)^2}\delta^2(\theta_1-\theta_2)\nonu
&&\times
\frac{\lambda m-D^2_{k\theta_2}}{k^2+(\lambda m)^2}\delta^2(\theta_2-\theta_3)
\frac{\lambda m-D^2_{p\theta_3}}{p^2+(\lambda m)^2}\delta^2(\theta_3-\theta)\nonumber
\eea

\nin Using again the properties (\ref{propgates}), we obtain

\bea\label{cub}
&&\frac{g^3}{(2\pi)^9}\int\frac{d^3p~d^3q~d^3k~d^2\theta}
{[p^2+(\lambda m)^2]^2[(p+q)^2+(\lambda m)^2][(p-k)^2+(\lambda m)^2]}\nonu
&&\times\left(4(\lambda m)^3 Q(k,\theta)Q(q,\theta)Q(-k-q,\theta)\right.\nonu
&&~~+6(\lambda m)^2 Q(q,\theta)Q(-k-q,\theta)D^2_{-p}Q(k,\theta)\nonu
&&~~+4\lambda m Q(-k-q,\theta)D^2_{-p-q}[Q(q,\theta)D^2_{-p}Q(k,\theta)]\nonu
&&~~\left.+p^2Q(q,\theta)Q(k,\theta)D^2_{k-p}Q(-k-q,\theta)\right)
\eea

\nin The evolution of the coupling constant $g$ is obtained in the limit of zero
incoming momenta, since in the approximation (\ref{GammaQ}) the derivative 
interactions are neglected. Considering the fact that

\be
D^2_{-p-q}D^2_{-p}=-p^2+D^2_{-q}D^2_{-p}-2pq\theta^2D^2_{-p},
\ee

\nin the non-derivative interaction terms coming from Eq.(\ref{cub}) are then

\bea
&&g^3\int\frac{d^3q}{(2\pi)^3}\frac{d^3k}{(2\pi)^3}d^2\theta
~~ Q(k,\theta)Q(q,\theta)Q(-k-q,\theta)\\
&&\times\int\frac{d^3p}{(2\pi)^3}\frac{4(\lambda m)[(\lambda m)^2-p^2]}
{[p^2+(\lambda m)^2]^2[(p+q)^2+(\lambda m)^2][(p-k)^2+(\lambda m)^2]},\nonumber
\eea

\nin for $k=q=0$ in the integrand. The identification of both sides
of the evolution equation (\ref{evolsusy1}) gives then Eq.(\ref{evolg}).

\vspace{1cm}

We finally compute the one-loop renormalization of this model, using the 
Feynman graphs technique, and starting with the bare action for $\lambda=1$:

\be
S_{\lambda=1}=\int d^5z \left\{\hf QD^2Q +\frac{m_0}{2}Q^2+\frac{g_0}{6}Q^3\right\}.
\ee

The one-loop correction to the quadratic term in $Q$ is given by the self-energy
graph 

\bea
\Gamma_2[Q]&=&g_0^2\int\frac{d^3p}{(2\pi)^3}\frac{d^3q}{(2\pi)^3}d^2\theta d^2\theta'
Q(-p,\theta)Q(p,\theta')\nonu
&&\times\frac{m_0-D^2_{q\theta}}{m_0^2+q^2}\delta^2(\theta-\theta')
\frac{m_0-D^2_{q+p\theta'}}{m_0^2+(q+p)^2}\delta^2(\theta'-\theta).
\eea

\nin Using the properties (\ref{propgates}), we obtain

\bea
\Gamma_2[Q]&=&g_0^2\int\frac{d^3p}{(2\pi)^3}\frac{d^3q}{(2\pi)^3}d^2\theta d^2\theta'
\frac{1}{(m_0^2+q^2)[m_0^2+(q+p)^2]}\\
&&\times\LARGE\{-2\delta^2(\theta-\theta')m_0Q(-p,\theta)Q(p,\theta)\nonu
&&~~+\delta^2(\theta-\theta')Q(p,\theta')D^2_{q\theta}[Q(-p,\theta)D^2_{q+p\theta'}
\delta^2(\theta'-\theta)]\LARGE\}\nonu
&=&g_0^2\int\frac{d^3p}{(2\pi)^3}\frac{d^3q}{(2\pi)^3}d^2\theta
\frac{Q(p,\theta)D^2_{q\theta}Q(-p,\theta)-2m_0Q(p,\theta)Q(-p,\theta)}
{(m_0^2+q^2)[m_0^2+(q+p)^2]}.\nonumber
\eea

The correction to the mass is obtained for $p=0$ in the non-derivative integrand, 
such that the one-loop mass is 

\bea
m^{(1)}&=&m_0-\hbar g_0^2\int\frac{d^3q}{(2\pi)^3}\frac{2m_0}{(m_0^2+q^2)^2}\nonu
&=&m_0-\hbar\frac{g_0^2}{4\pi},
\eea

\nin and the constant $c$ appearing in Eq.(\ref{mfinal}) is, at one loop, 
$c=-\hbar g_0^2/(4\pi m_0)$.

The one-loop correction to the cubic term in $Q$ is given by

\bea
\Gamma_3[Q]&=&g_0^3\int\frac{d^3p}{(2\pi)^3}\frac{d^3q}{(2\pi)^3}\frac{d^3k}{(2\pi)^3}
d^2\theta d^2\theta_1d^2\theta_2\\
&&\times Q(p,\theta)Q(q,\theta_1)Q(-p-q,\theta_3)\nonu
&&\times \frac{m_0-D^2_{k\theta}}{m_0^2+k^2}\delta^2(\theta-\theta_1)
\frac{m_0-D^2_{k+p\theta_1}}{m_0^2+(k+p)^2}\delta^2(\theta_1-\theta_2)\nonu
&&\times\frac{m_0-D^2_{k+p+q\theta_2}}{m_0^2+(k+p+q)^2}\delta^2(\theta_2-\theta)\nonumber.
\eea

\nin The properties (\ref{propgates}) give then

\bea
\Gamma_3[Q]&=&\frac{g_0^3}{(2\pi)^9}\int\frac{d^3p~d^3q~d^3k~d^2\theta}
{(m_0^2+k^2)[m_0^2+(k+p)^2][m_0^2+(k+p+q)^2]}\nonu
&&~~\times\LARGE\{3m_0Q(-p-q,\theta)D^2_{k+p\theta}[Q(p,\theta)Q(q,\theta)]\nonu
&&~~~~-3m_0^2Q(p,\theta)Q(q,\theta)Q(-p-q,\theta)\nonu
&&~~~~-D^2_{k+p\theta}[Q(q,\theta)Q(-p-q,\theta)D^2_{k\theta}Q(p,\theta)]\LARGE\}\nonu
&=&g_0^3\int\frac{d^3p}{(2\pi)^3}\frac{d^3q}{(2\pi)^3}\frac{d^3k}{(2\pi)^3}d^2\theta\nonu
&&\times\frac{(k^2-3m_0^2)Q(p,\theta)Q(q,\theta)Q(-p-q,\theta)}
{(m_0^2+k^2)[m_0^2+(k+p)^2][m_0^2+(k+p+q)^2]}\nonu
&&~~~~+\mbox{derivative terms}
\eea

\nin The correction to the coupling constant is obtained in the limit $p,q\to 0$ in the 
non-derivative integrand, i.e. it is given by the integral

\be
g_0^3\int\frac{d^3k}{(2\pi)^3}\frac{k^2-3m_0^2}{(m_0^2+k^2)^3}=0,
\ee

\nin such that the one-loop correction to the coupling constant vanishes, in accordance 
with the the result (\ref{finalg}).

\end{document}